\documentclass[11pt,twoside]{article}

\usepackage{sproclgross,epsfig}

\def\Journal#1#2#3#4{{#1} {\bf #2}, #3 (#4)}

\def\AP{\em Ann. Phys.}

\def\PLB{{\em Phys. Lett.}  B}
\def\PRL{\em Phys. Rev. Lett.}
\def\PRC{{\em Phys. Rev.} C}

\def\bra#1{\langle {#1} |\, }
\def\ket#1{\, | {#1} \rangle}
\def\braw#1{\langle \widehat{{#1}} |\, }
\def\ketw#1{\, | \widehat{{#1}} \rangle}
\def\erw#1{\,\langle \, {#1} \, \rangle\,}
\newcommand{\braket}[2]{\langle {#1} | {#2} \rangle}

\newcommand{\op}[1]{#1}
\newcommand{\coop}[1]{\widehat{#1}}
\newcommand{\vek}[1]{{\!\vec{\,#1}}}
\newcommand{\ddt}{\frac{\mathrm{d}}{{\mathrm{d}}t}\:}
\newcommand{\dint}{{\mathrm{d}}}
\newcommand{\im}{{\mathrm{i}}\, }

\newcommand{\pp}[2]{\frac{\partial  {#1}}{\partial  {#2}}\:}
\begin{document}

\title{FERMIONIC MOLECULAR DYNAMICS AND SHORT RANGE CORRELATIONS
\footnote{Invited talk at "XVII RCNP International Symposium on Innovative
Computational Methods in Nuclear Many-Body Problems",
November 10 - 15, 1997, Osaka. To be published in the proceedings, 
World Scientific.}}

\author{H. FELDMEIER, T. NEFF, R. ROTH}

\address{Gesellschaft f\"ur Schwerionenforschung mbH, D-64220 Darmstadt, 
Germany} 
\author{J. SCHNACK}
\address{Fachbereich Physik, Universit\"at Osnabr\"uck, D-49069 Osnabr\"uck}

\maketitle\abstracts{ 
Fermionic Molecular Dynamics (FMD) models a system of 
fermions by means of many--body states which are composed of antisymmetrized
products of single--particle states. These consist of one or several 
Gaussians localized  in coordinate and momentum space. The
parameters specifying them are the dynamical variables of the model. 
As the repulsive core of the nucleon--nucleon interaction induces short 
range correlations which cannot be accommodated by a Slater determinant, 
a novel approach, the unitary correlation operator method (UCOM), is applied.  
The unitary correlator moves two particles away from each other whenever their 
relative distance is within the repulsive core. 
The time--dependent variational principle yields the
equations of motion for the variables. 
Energies of the stationary ground states are calculated and compared 
to exact many--body results for nuclei up to $^{48}$Ca. 
Time--dependent solutions are shown for collisions between nuclei. 
}

%%%%%%%%%%%%%%%%%%%%%%%%%%%%%%%%%%%%%%%%%%%%%%%%%%%%%%%%%%%%%%%%%%%%%%%%
\section{Concept}

The general concept of Fermionic Molecular Dynamics (FMD) \cite{FMDpapers}
is to describe
a system of fermions by a many--body trial state which is antisymmetric 
with respect to particle exchange ("Fermionic"). The attribute
"Molecular" indicates that the individual single--particle states,
from which Slater determinants are formed, are 
wave packets localized in coordinate
as well as momentum space. Hence they are the quantum analogue to 
classical points in phase space. The time evolution of the parameters
which specify the many--body state, like mean positions, mean momenta, 
or spin orientations of the wave packets, is obtained from the 
time--dependent variational principle ("Dynamics"). 
A closely related model is AMD~\cite{AMDpapers} which follows the 
same concept but differs from FMD in several aspects.  

%%%%%%%%%%%%%%%%%%%%%%%%%%%%%%%%%%%%%%%%%%%%%%%%%%%%%%%%%%%%%%%%%%%%%%%%
\subsection{The trial state}
\label{TRIAL}

In FMD the trial states are antisymmetrized $A$--body states
%-------------------------------------------------------------------
\begin{eqnarray}\label{trial}
\ket{Q}=\op{C} \op{{\cal A}} \;\ket{\hat{q}_1} \otimes 
\ket{\hat{q}_2} \otimes \cdots \otimes \ket{\hat{q}_A} \ ,
\end{eqnarray}
%-------------------------------------------------------------------
where $Q$ denotes the set of all subsets $\hat{q}_k$ which specify 
the parameters of the single--particle states $\ket{\hat{q}_k}$. 
These are Gaussians, localized in 
phase space, or superpositions of several Gaussians. In this case $\hat{q}_k$ 
includes in addition to the complex 
parameters mean positions, widths and 
two--component spinors also the complex amplitude for each Gaussian.
In general $\ket{Q}$ can be a superposition of correlated Slater 
determinants, 
then the set $Q$ includes also the configuration--mixing amplitudes.

$\op{{\cal A}}$ is the projector onto the antisymmetric Hilbert space of
fermions. $\op{C}$ is a unitary correlation operator which takes care of 
the short range correlations as explained in subsection~\ref{UCOM}.

%%%%%%%%%%%%%%%%%%%%%%%%%%%%%%%%%%%%%%%%%%%%%%%%%%%%%%%%%%%%%%%%%%%%%%%%
\subsection{The time--dependent variational principle}
\label{TDVP}

The equations of motion for all parameters in 
$Q=\{q_{\nu};\nu=1,2,\cdots,N\}$ are obtained from the time--dependent 
variational principle
%------------------------------------------------------------------
\begin{eqnarray}\label{tdvp}
\delta \int_{t_1}^{t_2} \! \! \dint t \;
\bra{Q(t)}\; i \ddt - \op{H} \; \ket{Q(t)}
\ =\ 0 \ ,
\end{eqnarray}
%------------------------------------------------------------------
where the variation is taken with respect to all individual parameters
$q_\mu$ or their 
complex conjugate $q_\mu^*$. The resulting equtions of motion are
%------------------------------------------------------------------
\begin{eqnarray}\label{eom}
\bra{\pp{}{q_\mu^*} Q(t)}
 \; \im \ddt - \op{H} \; \ket{Q(t)} = 
\bra{\pp{}{q_\mu^*} Q(t)}\im \! \! \sum_\nu 
           \dot{q}_\nu \pp{}{q_\nu} - \op{H} \; \ket{Q(t)}\ =\ 0 \ .
\end{eqnarray}
%------------------------------------------------------------------
The physical meaning of these equations is that the deviation of
the exact state from the approximate $\ket{Q(t+\dint t)}$,
which develops during a time interval d$t$,
%------------------------------------------------------------------
\begin{eqnarray}
\ket{Error(t+\dint t)} &\equiv&
 \; \big(\: 1 -\im \dint t \;\op{H} \: \big) \ket{Q(t)} - \ket{Q(t+\dint t)} \\
&=& 
\im \dint t \big( \; \im \! \! \sum_\nu 
           \dot{q}_\nu \pp{}{q_\nu} - \op{H} \;\big) \ket{Q(t)}\label{Error}
\end{eqnarray}
%------------------------------------------------------------------
is orthogonal to the tangent 
$\bra{\pp{}{q_\mu^*} Q(t)}^{\dagger}=\ket{\pp{}{q_\mu}Q(t)}
\equiv \pp{}{q_\mu}\!\ket{Q(t)}$
at each time $t$, as can be seen by 
comparing Eq.~(\ref{Error}) with (\ref{eom}). 
In other words the variational principle chooses 
the best $\ket{Q(t+\dint t)}$ within the manifold $\ket{Q}$.

Stationary solutions of the variational principle (\ref{tdvp}), 
ground states for example, imply
%------------------------------------------------------------------
\begin{eqnarray}\label{stat}
\bra{\pp{}{q_\mu^*} Q(t)}
 \; \op{H} \; \ket{Q(t)} = 0 \ .
\end{eqnarray}
%------------------------------------------------------------------

%%%%%%%%%%%%%%%%%%%%%%%%%%%%%%%%%%%%%%%%%%%%%%%%%%%%%%%%%%%%%%%%%%%%%%%%
\subsection{Quantum--branching}
\label{qu-branch}
If one wants to allow the time evolution to deviate from $\ket{Q(t)}$
without increasing the degrees of freedom one may take the operator
%------------------------------------------------------------------
\begin{eqnarray}\label{pertubation}
\op{\Delta V(t)}\equiv \big( \; \im \! \! \sum_\nu 
           \dot{q}_\nu \pp{}{q_\nu} - \op{H} \; \big)
\end{eqnarray}
%------------------------------------------------------------------
from Eq.~(\ref{Error})
as a time--dependent pertubation to calculate transition amplitudes
for quantum--branching away from $\ket{Q(t)}$. 
If the time evolution within the manifold
happens to be exact, $\op{\Delta V(t)}\ket{Q(t)}=0$ and no branching
occurs because all quantum physics is already in $\ket{Q(t)}$.
The branching introduced by Ono and Horiuchi
in AMD-V~\cite{AMD-V} or the branching between energy eigenstates 
proposed by Onishi and Randrup \cite{OniRan} may be investigated 
under this aspect.

%%%%%%%%%%%%%%%%%%%%%%%%%%%%%%%%%%%%%%%%%%%%%%%%%%%%%%%%%%%%%%%%%%%%%%%%
\section{The Unitary Correlation Operator Method (UCOM)}
\label{UCOM}

In nuclear physics this straightforward concept encounters a 
complicated Hamiltonian. Due to the complex structure of the 
nucleons and the mesons exchanged between them, 
the interaction depends on spin and 
isospin, is momemtum dependent and overall repulsive at short distances,
see for example Ref.~\cite{BonnPot}. The properties of the Hamiltonian 
are of course intimately related to the choice of the trial state. 
The tensor part induces a strong correlation between the spins of a pair 
of nucleons and the direction of their relative 
distance~\cite{Pandharipande}. 
The repulsion at short distances reduces the probability amplitude to 
find two nucleons close together. Both correlations cannot be described 
by a (antisymmetrized) product state. Therefore the most simple ansatz
of a single Slater determinant has to be modified in order
to incorporate the above mentioned correlations.
In the following we sketch the new Unitary Correlation Operator
Method (UCOM) which allows an approximate treatment of the short range 
repulsion even in time--dependent states.

Due to the short--ranged repulsive core in the nucleon--nucleon 
potential $V$ the many--body state is depleted as a function of
the relative distance $x_{ij} = | \vec{x}_i - \vec{x}_j |$ for
each pair $(ij)$ when they are close to each other.
These short range correlations cannot be described by shell model
states.
The most common procedure to remedy this problem is Brueckner's
G-matrix method which replaces the bare $\op{V}$ by an effective
interaction $\op{G}$.
Another method is the Jastrow approach where the correlated
ground state of the nucleus is assumed to be of the form
$\prod_{i<j} f(x_{ij}) \ket{\Phi}$ where $f(x_{ij})$ is a
correlation factor which diminishes the probability to find
two nucleons at small distances $x_{ij}$.
We propose a new method \cite{UCOM}
in which the correlated state $\ket{Q}$ is obtained by a 
unitary transformation $\op{C}=\exp\{-\im \op{G}\}$. 
With respect to the choice of $\op{G}$
it differs from the work of K.~Suzuki et al.\cite{Suzuki}, 
where an effective interaction for the shell model is defined.
In collisions between nuclei a preferred shell model space does not
exist. Therefore, our definition of $\op{G}$ is not related to a given basis
but to the behaviour of the relative wave function at small distances.
The correlated state is given by  
%--------------------------------------------------------------
\begin{equation}
\ket{Q} = \op{C} \ketw{Q} = {\mathrm{e}}^{-\im \op{G}} \ketw{Q} \;,
\end{equation}
%--------------------------------------------------------------
where the generator $\op{G}$ is a hermitean two--body operator which depends
in general on the relative distance $\vek{\op{x}}_{ij}$, the relative
momentum  $\vek{\op{q}}_{ij}$, the spins and isospins 
of the two nucleons \cite{UCOM}. For sake of simplicity we sketch here only 
the case of a repulsive core without spin--isospin--dependence. 
In the application shown later $\op{G}$ depends on spin and 
isospin such that the different channels of a central interaction
can be correlated individually. 
 
The aim of $\op{C}$ is to push two nucleons away from each
other whenever they get too close.
The most simple ansatz which does that is
%--------------------------------------------------------------
\begin{equation}
\op{G} = \frac{1}{2}  \sum_{i<j} \left\{
\left( \vek{\op{q}}_{ij} \frac{\vek{\op{x}}_{ij}}{\op{x}_{ij}}\right) 
s(\op{x}_{ij}) + s(\op{x}_{ij})
\left( \frac{\vek{\op{x}}_{ij}}{\op{x}_{ij}}\vek{\op{q}}_{ij} \right)
\right\}
\ , \ \op{x}_{ij}\equiv | \vek{\op{x}}_{ij}| \; .
\end{equation}
%--------------------------------------------------------------
$s(x_{ij})$ is roughly speaking the  distance which the
particles $i$ and $j$ are moved away from each other by $\exp \{-\im \op{G}\}$ 
if they are found at a distance $x_{ij}$ in $\ketw{Q}$.
$s(x_{ij})$ is largest if $x_{ij}$ lies inside the hard core and 
$s(x_{ij}) \rightarrow 0$ if $x_{ij}$ is outside the repulsive
interaction.

As $\op{C}^\dagger=\op{C}^{-1}$ is unitary one can correlate the states
or equivalently the observables:
%--------------------------------------------------------------
\begin{equation} 
\bra{Q} \op{H} \ket{Q}
  = \braw{Q} \op{C}^{\dagger} \op{H} \,  \op{C} \ketw{Q}
 = \braw{Q} \op{C}^{-1} \op{H} \,  \op{C} \ketw{Q}
 \equiv \braw{Q}  \coop{H}    \ketw{Q} \, .
\end{equation}
%--------------------------------------------------------------

Closed expressions for the correlated relative wave function, 
the correlated potential and the correlated kinetic energy can be 
given by using the coordinate transformations
$x_{ij} \rightarrow  R_{\pm}(x_{ij})$ defined by
%--------------------------------------------------------------
\begin{equation}
\int_{x}^{R_{\pm}(x)} \frac{\dint \xi}{s(\xi)} = \pm 1 \; .
\end{equation}
%--------------------------------------------------------------
For example the correlated potential is simply
%--------------------------------------------------------------
\begin{equation}\label{coopV}
\coop{V}^{[2]}(x_{ij}) \equiv \op{C}^{-1} \op{V}(x_{ij}) \, \op{C} =
\op{V}(\op{C}^{-1}x_{ij} \, \op{C}) = V(R_+(x_{ij})) \ .
\end{equation}
%--------------------------------------------------------------
Or a correlated relative wave--function can be expressed in terms of 
$R_-$ which is the inverse of $R_+$ as
$(\vek{x} \equiv \vek{x}_i-\vek{x}_j = \vek{x}_{ij})$
%-------------------------------------------------------------------------
\begin{eqnarray}\label{E-2-2-G}
\braket{\vek{x}}{\phi} =
\bra{\vek{x}}\op{C}\ketw{\phi}
&=&
\frac{R_-(x)}{x} \sqrt{\frac{\mathrm{d}}{{\mathrm{d}}x}R_-(x)}\;\;
\braket{R_-(x)\frac{\vek{x}}{x}}{\widehat{\phi}}
\end{eqnarray}
%-------------------------------------------------------------------------

%===================    figure   =================================
\begin{figure}[!bt]
\begin{center}
\epsfig{file=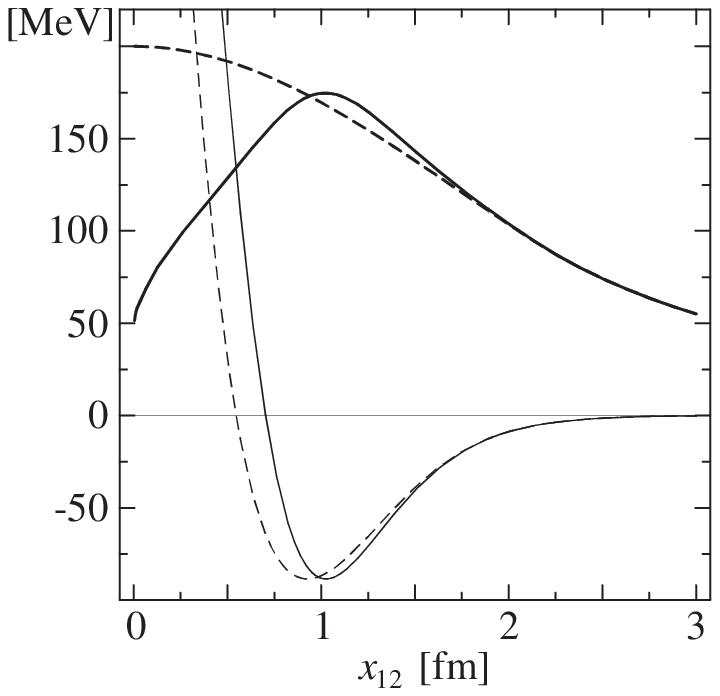,height=60mm}
\end{center}
\caption{Uncorrelated state 
$\braket{\vek{x}_{12}}{\widehat{\phi}}$(dashed line),
correlated state $\braket{\vek{x}_{12}}{\phi}$ (full line), 
Afnan--Tang potential $V(| \vek{x}_{12} |)$ (thin line) and 
correlated potential $\coop{V}^{[2]}(| \vek{x}_{12} |)$ (thin dashed line) 
as a function of the relative distance $x_{12}=| \vek{x}_{12} |$.}
\end{figure} 
%===================    figure   =================================

Fig. 1 displays the radial dependence of the correlated and
uncorrelated deuteron wave function together with the Afnan Tang S3
potential \cite{Afnan}.
The functional form of $s(x)$ or equvalently of $R_+(x)$ is chosen 
such that the correlated state $\ket{\phi}$ for short distances 
equals the exact solution. The correlated kinetic energy is more involved
as one has to calculate the correlated relative momentum 
$\op{C}^{-1} \op{\vek{q}}_{ij} \, \op{C}$. 
The details are given in Ref.~\cite{UCOM}.

Once the correlator $\op{C}$ is adjusted to reproduce the
two--body system at low energies (long wave length) one can calculate
the ground state energies of nuclei by mini\-mizing
$\bra{Q} \op{H} \ket{Q} = \braw{Q}  \coop{H}    \ketw{Q}$
with respect to $\ketw{Q}$ which in this application is a single 
Slater determinant composed of localized Gaussians (FMD).

Unlike the original Hamiltonian $\op{H} = \op{T} + \op{V}$, the 
correlated Hamiltonian
%-------------------------------------------------------------------------
\begin{eqnarray}\label{expansion}
\hspace*{-3mm}
\coop{H} \equiv   \op{C}^{-1} H \, \op{C} =  
\op{C}^{-1}\op{T}\, \op{C}  + \op{C}^{-1} \op{V} \,  \op{C} 
= \op{T} + \!\coop{T}^{[2]}+ \! \coop{V}^{[2]} + \!
\coop{T}^{[3]} + \! \coop{V}^{[3]} + \cdots
\end{eqnarray}
%-------------------------------------------------------------------------
is not a one-- plus two--body operator anymore. It contains three--body 
and higher interactions because the generator $\op{G}$ is a two--body operator.

We calculate $\coop{T}^{[2]}$ and $\coop{V}^{[2]}$, 
which are functionals of $R_+$, analytically 
(for example as in Eq.~(\ref{coopV})) and
neglect irreducible three--body and higher terms in 
the expansion (\ref{expansion}). 
This turns out to be a very good approximation at typical
nuclear densities. Estimations of the three--body terms give
corrections less than 5\% of binding energy for the
$\alpha$-particle \cite{UCOM}. 
%===================    figure   =================================
\begin{figure}[!bt]
\begin{center}
\epsfig{file=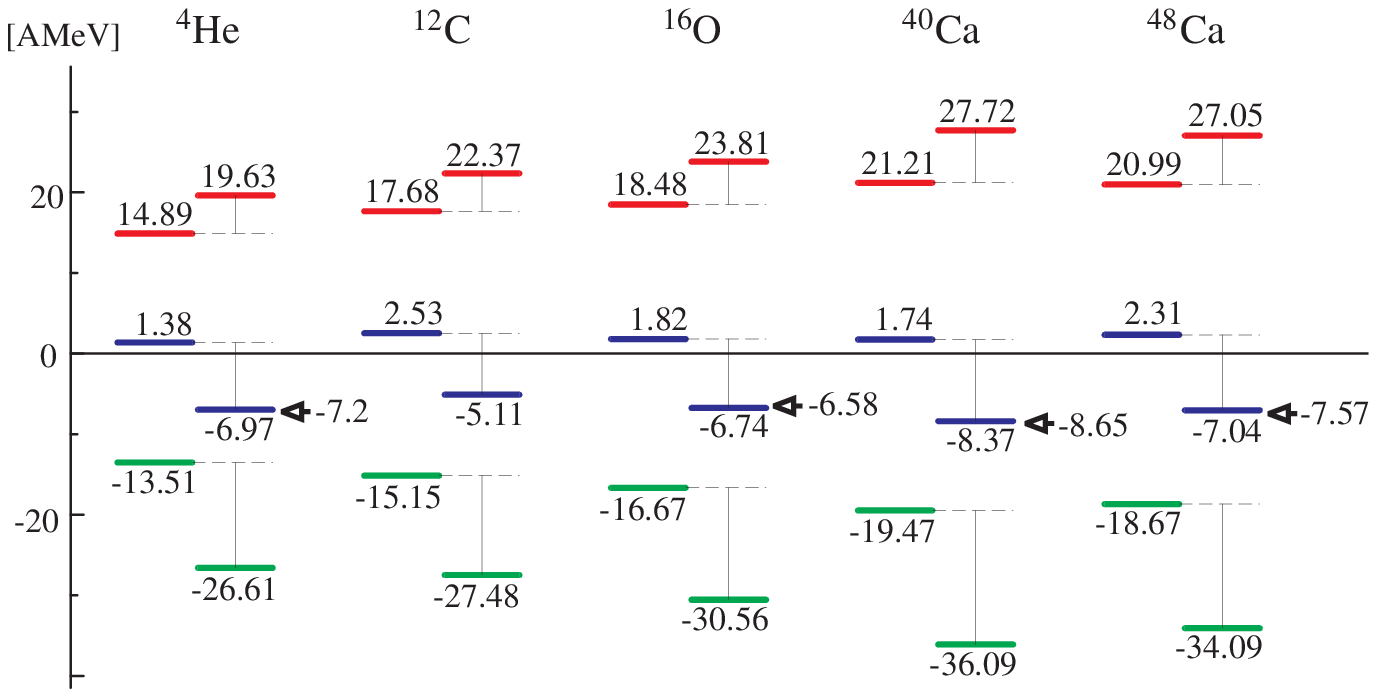,width=160mm}
\end{center}
\caption{
Energies for the modified Afnan-Tang S3 potential.
Left columns: expectation values of kinetic 
energy $\erw{\op{T}}$, potential $\erw{\op{V}}$ and 
sum of both $\erw{\op{H}}$. Right columns: expectation
values of correlated kinetic energy $\erw{\op{T}+\coop{T}^{[2]}}$, 
correlated potential $\erw{\coop{V}^{[2]}}$ and sum of both
$\erw{\coop{H}^{C2}}$.
Arrows indicate the results of a
Yakubovski calculation for $^4$He and  
of CBF calculations for $^{16}$O and 
for $^{40,48}$Ca.       
}
\end{figure} 
%===================    figure   =================================

Fig. 2 compares uncorrelated (left columns) and correlated (right
columns) energies.
The correlated potential energy $\braw{Q}\coop{V}^{[2]} \ketw{Q}$ 
(grey bars at negative values) is about twice the uncorrelated 
$\braw{Q} \op{V} \ketw{Q}$ in all nuclei.
This gain in binding is counteracted by an increase in the
kinetic energies (dark grey bars at positive values).
Both together yield binding energies (black bars) which are
within 8\% deviation from results of Yakubovski calculations \cite{KaG92}
for $\mathrm{{}^4He}$ and CBF calculations \cite{CFF92,SCF96} for
$\mathrm{{}^{16}O}$ and $\mathrm{{}^{40,48}Ca}$. 
This suprises since $\coop{H}^{C2}=\op{T}+\coop{T}^{[2]}+\coop{V}^{[2]}$ 
is the same for all nuclei and not density dependent but it contains 
momentum dependent parts in the two--body part of the correlated kinetic 
energy $\coop{T}^{[2]}$. 

%%%%%%%%%%%%%%%%%%%%%%%%%%%%%%%%%%%%%%%%%%%%%%%%%%%%%%%%%%%%%%%%%%%%%%%%
\section{Ground states with several Gaussians}
\label{halo}
The Gaussian single--particle states do not describe properly the 
surface of nuclei because their tails do not fall off exponentially.
Especially in the case of halo nuclei this is an important part of the 
physics. As Gaussians represent an overcomplete basis, any shape can 
be achieved by superimposing several Gaussians. In the following example 
we minimize the energy of $^6$He for a single Slater determinant 
with one Gaussian per particle and with two co--centered Gaussians with 
free widths and relative amplitudes. The resulting proton and neutron  
densities are displayed in Fig.~3 on a logarithmic scale. 
For the one Gaussian case (l.h.s.) the distribution of the neutrons 
is, as expected, wider than for the protons, but already with two Gaussians 
(r.h.s.) the trial state is much improved and exhibits nice exponential 
tails of different ranges. 
The binding energy increases 
by about 3.5 MeV and the $^4$He core which is distorted on the 
l.h.s. assumes again equal proton and neutron densities in the center. 
Three Gaussians do not improve the picture anymore. This example shows that
a Gaussian basis may be a good representation for nuclear structure 
calculations. See for example also the contribution by M. Kamimura 
in these proceedings.  

%===================    figure   =================================
\begin{figure}[bt]
\begin{center}
\epsfig{file=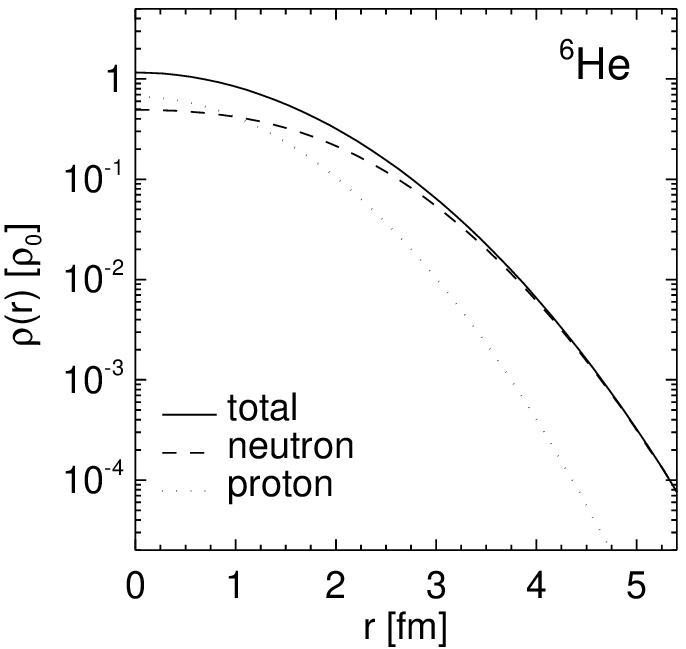,height=65mm}\hspace*{10mm}
\epsfig{file=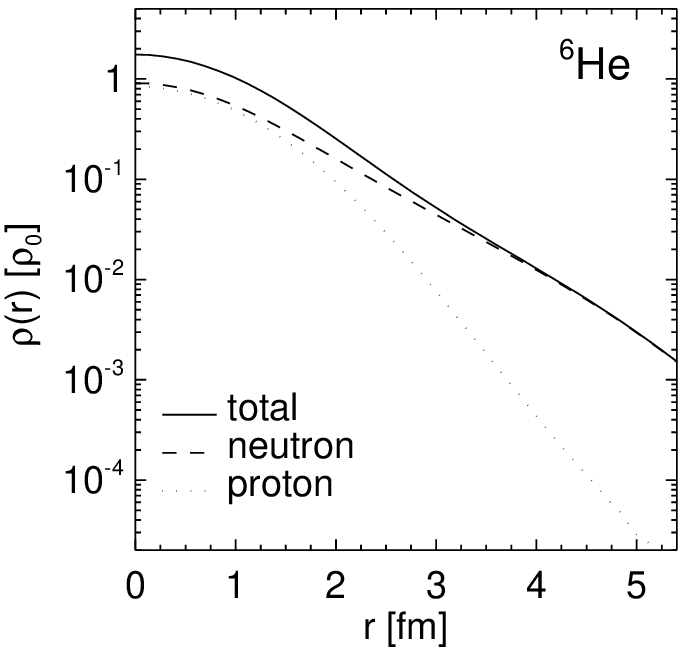,height=65mm}

\end{center}
\caption{
Proton and neutron densities of $^6$He with one Gaussian per aprticle
(l.h.s.) and two Gaussians per particle (r.h.s)       
}
\end{figure} 
%===================    figure   =================================

We have also diagonalized the Hamilton matrix 
$\bra{Q_k}\op{H}\ket{Q_l}$ within a subspace spanned by the nonorthogonal
many--body states $\ket{Q_k}$. Dimensions of up to several hundred
are feasible and for example eigenstates of total spin may be obtained
by taking  $\ket{Q_k}$ as rotated intrinsic states. See also contributions 
of S. Aoyama, N. Itagaki and S. Okabe or Y. Kanada-En'yo and H. Horiuchi
or of K. Varga and Y. Suzuki in these proceedings.

%%%%%%%%%%%%%%%%%%%%%%%%%%%%%%%%%%%%%%%%%%%%%%%%%%%%%%%%%%%%%%%%%%%%%%%%
\section{The time--dependent case}
\label{CaCa}

In the time--dependent case one cannot be so ambitious as for ground 
states because all derivatives in Eq.~\ref{eom} have to be calculated 
at each time step. Here we have to compromise and take only one Gaussian 
per particle and one Slater determinant. In addition the Hamiltonian 
is a simplified version of $\coop{H}=\op{C}^{-1}\op{H}\op{C}$. 
We tried to adjust the parameters of the central $\coop{H}$ such 
that ground state energies of nuclei up to mass 56 are reproduced 
within about 5\% error. In doing that we found several Hamiltonians
which describe similarly well the ground states but differ in their 
momentum dependence. A particular $\coop{H}$ has two almost 
degenerate local energy minima in the manifold $\ket{Q}$. 
One minimum has states with an intrinsic cluster 
structure with narrow Gaussians placed at different positions in
coordinate space, but centered in momentum space. The other shows 
co--centered Gaussians in cordinate space which are slightly displaced 
in momentum space, which resembles a shell model state.   
%===================    figure   =================================
\begin{figure}[!ht]\label{Ca}
\begin{center}
\epsfig{file=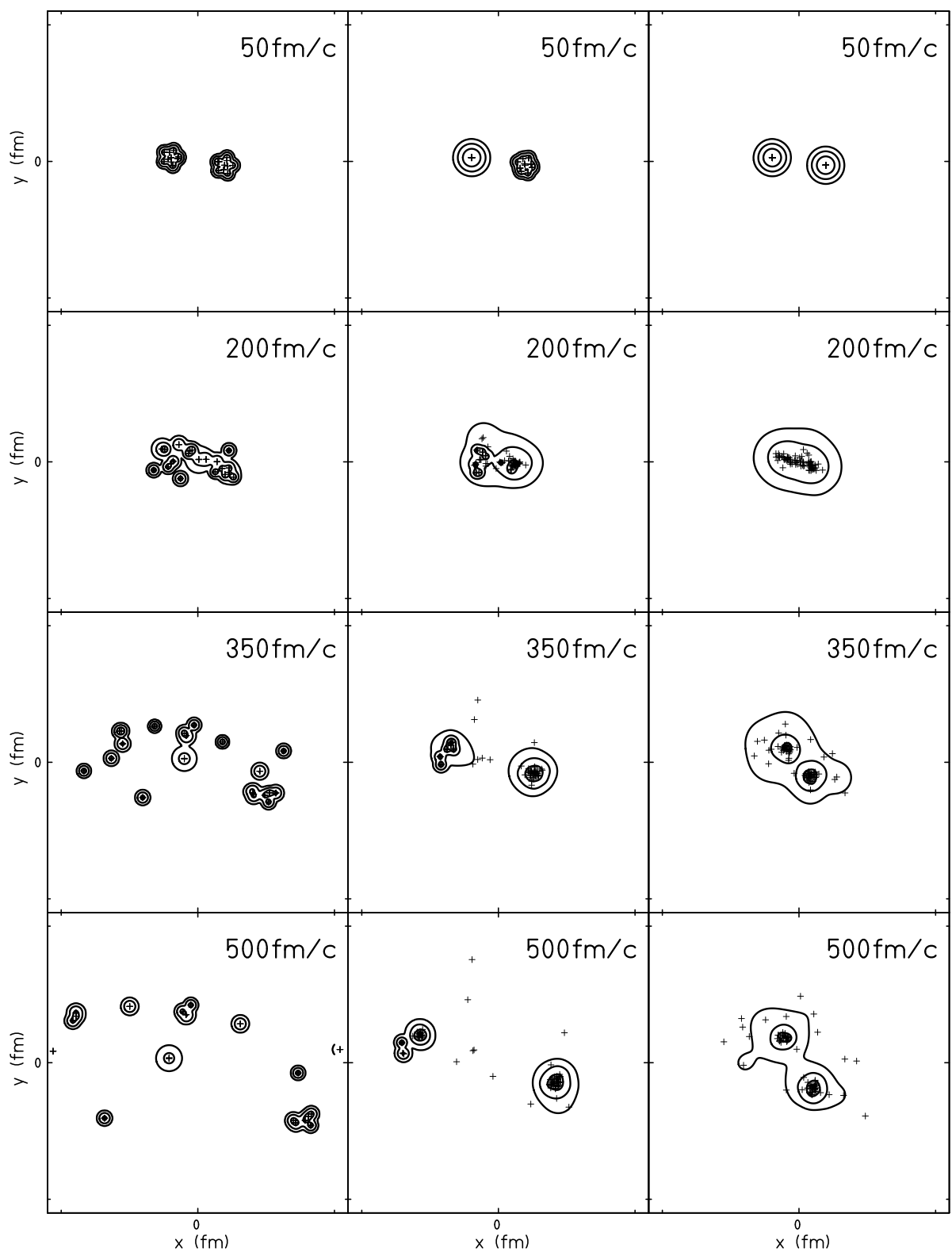,width=160mm}
\end{center}
\caption{
Densities projected on scattering plane for $^{40}$Ca+$^{40}$Ca collisions 
at b=2.75 fm and $E_{lab}$=35 AMeV. Initial nuclei in cluster and shell--model 
configurations. Crosses denote centroids of wave packets.}
\end{figure} 
%===================    figure   =================================
In Fig.~\ref{Ca} one sees that the two different types of states
behave quite differently when they collide. In the left column, 
where two cluster states interact, multifragmentation is observed 
which could be characterized as shattering of the nuclei. 
Shell model states on cluster states produce also several fragments 
but in addition many nucleons are emitted. They are seen in the 
middle column as crosses (centroids of the packets) without 
surrounding contour lines, because the widths are already so 
large that the density distribution of the emitted nucleon has 
fallen below the lowest contour line. Finally, reactions with 
two shell model 
states (right column) look more like a deep inelastic collision 
with many nucleons evaporated because of the high excitation energy.

%%%%%%%%%%%%%%%%%%%%%%%%%%%%%%%%%%%%%%%%%%%%%%%%%%%%%%%%%%%%%%%%%%%%%%%%%%%%

\end{document}